\documentclass{elsart}

\journal{Solid State Commun.}
\date{6 December 2006}

\usepackage{graphicx}

\usepackage{amssymb}
\usepackage{amsmath}
\usepackage[dvips]{color}

\begin{document}

\begin{frontmatter}

\title{The solid state phase transformation of potassium sulfate}

\author[1]{S. Bin Anooz}\ead{s\underline~binanooz@yahoo.com},
\author[2]{R. Bertram}\ead{bertram@ikz-berlin.de} \and
\author[2]{D. Klimm\corauthref{cor1}}\ead{klimm@ikz-berlin.de}
\corauth[cor1]{corresponding author}

\address[1]{Physics Department, Faculty of Science, Hadhramout University of Science \& Technology, Mukalla 50511, Republic of Yemen}
\address[2]{Institute for Crystal Growth, Max-Born-Str. 2, 12489 Berlin, Germany}

\begin{abstract}
Potassium sulfate single crystals that are grown from aqueous solutions lose upon the first heating up to 1\% of mass that is assumed to be water. This mass loss occurs in the vicinity of the PT from orthorhombic to hexagonal K$_2$SO$_4$. Only in the first heating run of K$_2$SO$_4$ that has not yet released water, pretransitional thermal effects can be observed in the DTA curve. If K$_2$SO$_4$ crystals are grown from solutions containing 4\,wt.\% Cd, Cu, or Fe, only Cu or Fe can be incorporated significantly with concentrations of several 0.1\%. The phase transformation temperature measured for such solid solutions depends on the heating rate. For pure K$_2$SO$_4$, the phase transformation temperature is independent on heating rate $581.3^{\,\circ}$C and the enthalpy of transformation is $(5.8\pm0.2)$\;kJ/mol.
\end{abstract}


\end{frontmatter}

\section{Introduction}

Potassium sulfate K$_2$SO$_4$ belongs at room temperature $T=25^{\,\circ}$C to the orthorhombic system and has four formula units per $D^{16}_{2h} = Pnma$ (olivine type) unit cell, with lattice parameters $a_0 = 7.476$\;\AA, $b_0 = 10.071$\;\AA, and $c_0 = 5.763$\;\AA\  \cite{Mcginnety72}. The substance transforms upon heating at $T_\text{t}\approx587^{\,\circ}$C into a hexagonal structure $D^{4}_{6h} = P6_3/mmc$ with $a_0 = 5.92$\;\AA\ and $c_0 = 8.182$\;\AA\ (measured at $640^{\,\circ}$C) where the oxygen positions of the SO$_4^{2-}$ tetrahedra are only partially occupied \cite{Arnold81}. In the following, the hexagonal high $T$ phase will be called $\alpha$-K$_2$SO$_4$ and the orthorhombic phase will be called $\beta$-K$_2$SO$_4$. Unfortunately, the opposite denomination is sometimes used in the literature \cite{Arnold81}. An analogous phase transformation (PT) to a $\alpha$-K$_2$SO$_4$ type phase at high $T$ is shown by other K$_2$SO$_4$-family crystals (e.g. Na$_2$SO$_4$, LiKSO$_4$, K$_2$CrO$_4$, and K$_2$SeO$_4$), in spite of structural differences at lower $T$ \cite{Aleksandrov91,Russell86}. Another PT of second order at 56\;K was derived from the temperature dependence of the lattice parameters. The crystal symmetry of this low-temperature phase ($\gamma$-K$_2$SO$_4$) is assumed to be monoclinic \cite{Ahmed96}.

Other authors observed the $\alpha-\beta$ transformation of K$_2$SO$_4$ at $592^{\,\circ}$C or $612^{\,\circ}$C, respectively \cite{Barbooti77,Drona82}, but these investigations were performed with mixtures instead of pure K$_2$SO$_4$. Electrical conductivity measurements on single crystals have been carried out by Choi et al. \cite{Choi93}, the transition temperature range extends to 3\;K and transition occurs at $586.9^{\,\circ}$C on heating and $581.5^{\,\circ}$C on cooling, with a thermal hysteresis of 5.4\;K. Chen et al. performed electrical complex impedance measurements on K$_2$SO$_4$ single crystals \cite{Chen00x}. The $T$ dependence of electrical conductivity and dielectric constant revealed an anomaly around $587^{\,\circ}$C which was attributed to the transformation to $\alpha$-K$_2$SO$_4$. Temperature dependence of Raman spectra of K$_2$SO$_4$ was measured from 27 to $900^{\,\circ}$C \cite{Ishigame83}. A remarkable change of the Raman frequencies is not observed through the phase transition temperature $T_\text{t}$.

Choi et al. \cite{Choi93} observed for a heating rate of 1\;K/min that most of the samples crack near $500\pm30^{\,\circ}$C showing an abrupt drop in electrical conductivity. In the heating process, elastic distortions, caused by the rotational disorder of sulfate ions, may be produced inside the crystal. Hence it is suggested that the occurrence of cracks around $500^{\,\circ}$C is further minor evidence for the existence of a pretransitional region in advance of the structural transition. Diosa et al. \cite{Diosa05} reported recently a deviation from the Arrhenius behavior of the complex dielectric permittivity of polycristalline K$_2$SO$_4$ in the temperature range $538-587^{\,\circ}$C, which is attributed to the onset of disordering of SO$_4^{2-}$ tetrahedra and a `pretransition' phenomenon as discussed by Choi et al. \cite{Choi93}. Diosa et al. \cite{Diosa05} did not comment on the remarkable difference between the dc-conductivity vs. $T$ curve obtained during heating of fresh material in comparison to subsequent heating/cooling runs (Fig.~1 of their paper).

The theoretical entropy change calculated from Boltzman's relation gives $\it\Delta$$S_\text{t} = 5.77$\;J/(mol\,K), provided that the configuration of the SO$_4^{2-}$ tetrahedron in $\alpha$-K$_2$SO$_4$ has two orientational possibilities. The observed entropy change, $\it\Delta$$S_\text{t} = 5.02$\;J/(mol\,K), agrees well with the calculation and the crystal structure of $\alpha$-K$_2$SO$_4$ appears reasonable in respect of the entropy change \cite{Miyake81}.

In literature one can observe differences in the reported data about high $T$ phase transition of K$_2$SO$_4$ crystal. Therefore we report here more detailed measurements of the PT using thermogravimetry (TG) and differential thermal analysis (DTA) techniques. Special emphasis is placed on the influence of water traces in the crystals grown from aqueous solution on pretransition phenomena. The investigation of doping effects with Cu$^{2+}$, Fe$^{2+}$, or Cd$^{2+}$ in small concentrations on the measured phase transition is another target of this article.

\section{Experimental}

Transparent and colorless crystals with well-defined edges (length up to 15\;mm) were obtained by the slow evaporation method of saturated aqueous K$_2$SO$_4$ solution. K$_2$SO$_4$ doped with Cu$^{2+}$, Fe$^{2+}$, or Cd$^{2+}$ was grown by the same method from solutions containing 4\,wt\% of the corresponding transition metal sulfate. Potassium sulfate usually crystallized in prismatic crystals, the $\vec{b}$-axis was found along the long axis of the prism and the $\vec{c}$-axis was along one edge of the quasi-triangular basal plane.

ICP-OES measurements compare the intensity of spectral lines obtained from the sample with the intensity of lines from calibrated standards. After careful calibration, main components as well as trace impurities down to 1\;ppm can be measured quantitatively with high precision. For the present study, an ``IRIS Intrepid HR Duo" (Thermo Elemental, USA) was used. The precision is $\approx3$\% relative standard deviation (R.S.D.) for concentrations above background equivalent concentration (BEC). For the undoped K$_2$SO$_4$ crystals only impurities on the ppm level could be found. The doped crystals contained the following concentrations of the dopant (in weight-\%): Cu$^{2+}$: = 0.192\%, Fe$^{2+}$ = 0.076\%, Cd$^{2+}$= 0.014\%.

X-ray measurement were performed with a single crystal powder diffractometer (Seifert URD6) using Cu K$_\alpha$ radiation. The K$_\beta$ line was suppressed with a thin Ni filter, whereas the Cu K$\alpha_1$--K$\alpha_2$ double peak structure was deconvoluted by using the Rachinger correction \cite{Rachinger48}. The measurements were swapped from $10^\circ \leq 2\,\theta \leq 90^\circ$ with a step width of $\it\Delta$\,$2\,\theta = 0.01^\circ$ and a scanning speed of about $0.003^\circ$/sec.

Thermal analysis was performed with a NETZSCH STA 449C ``Jupiter". A DTA sample carrier with each 2 series connected Pt90Rh10/Pt thermocouples for sample ($\approx10$\;mg) and reference was used with Al$_2$O$_3$ crucibles that were covered by a lid. All measurements were performed in flowing argon (99.999\% purity, 30\;ml/min). No calibration files were used for the measurements, as separate calibration measurements under identical conditions were performed with melting zinc ($T_\text{f} = 419.6^{\,\circ}$C, $\it\Delta$$H_\text{f} = 7322$\;J/mol) and with the first solid PT of barium carbonate BaCO$_3$ ($T_\text{t} = 805.8^{\,\circ}$C, $\it\Delta$$H_\text{t} = 18828$\;J/mol). These calibration values were taken from the FactSage 5.3. database \cite{FactSage5_4_1} were for the $\alpha-\beta$ transformation of K$_2$SO$_4$ one can find $T_\text{t} = 582.8^{\,\circ}$C, $\it\Delta$$H_\text{t} = 8954$\;J/mol. For the measurements heating/cooling rates from $\pm20$\;K/min down to $\pm1$\;K/min were used to check kinetic features.

The temperature $T$ during a DTA measurement is determined near the reference and the DTA signal is the difference between the sample temperature and $T$, hence an exothermal effect in the sample leads to a positive DTA signal. If a sudden heat pulse is produced in the sample the signal DTA$(t)$ ($t$ -- time) rises quickly to a maximum (peak, time $t_0$) and drops then exponentially
\begin{equation}
\text{DTA}(t) = k_1 \exp \left( \frac{-(t-t_0)}{\tau} \right) + k_2
\label{eq:exp}
\end{equation}
where the constants $k_1$ and $k_2$ describe the peak height and the position of the basis line and the time constant $\tau$ describes the thermal relaxation rate of the signal \cite{Hohne96}. It turned out, that the PT of the calibration substance BaCO$_3$ showed strong supercooling (up to 50\;K) if $T$ was lowered during DTA runs, resulting in a sharp peak rise near $T \approx 760^{\,\circ}$C. The dropping wing of the peak could very good be fitted to an exponential function (\ref{eq:exp}) giving $\tau = 49$\;s for the time constant of the DTA set-up that will be used in the following discussion.

\section{Results and Discussion}

\begin{figure}[htb]
\includegraphics[width=0.60\textwidth]{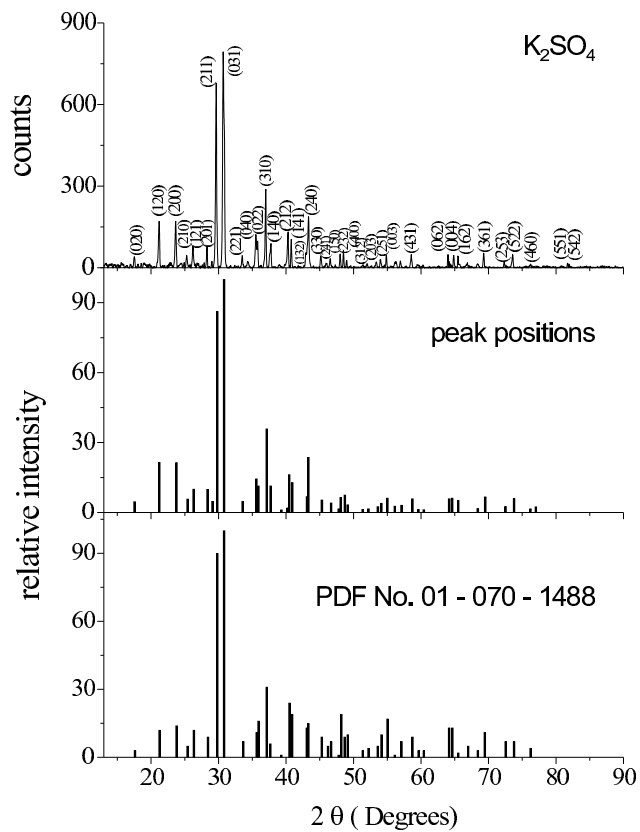}
\caption{Top: X-ray diffraction spectrum of K$_2$SO$_4$ crystal. Middle: Peak positions and relative intensity of the spectrum. Bottom: PDF No. 01-070-1488 ($\beta$-K$_2$SO$_4$) \cite{Mcginnety72}}
\label{fig:X-ray}
\end{figure}

The grown crystal of K$_2$SO$_4$ was crushed and subjected to powder X-ray diffraction analysis. A powder X-ray diffraction pattern of undoped potassium sulfate is shown in the top panel of Fig. \ref{fig:X-ray}. The diffraction peaks match very well the reported values of the peaks for K$_2$SO$_4$ crystal in the literature \cite{Mcginnety72}.
The data were treated by XPowder computer software \cite{XPowder04} and the calculated lattice parameters are in good agreement with the reported values for undoped crystal (Tab. \ref{tab:X-ray}).
\begin{table}[htbp]
\caption{Comparison of measured cell parameters with literature data \cite{Mcginnety72}. ($V$ -- unit cell volume)}
\begin{tabular}{lrrrr}
\hline
            & $a_0$ [\AA] & $b_0$ [\AA] & $c_0$ [\AA] & $V$ [\AA$^3$]\\
\hline
this work       & 7.489       & 10.073      & 5.763       & 434.7\\
PDF 01-070-1488 & 7.476       & 10.071      & 5.763       & 433.9\\
\hline
\end{tabular}
\label{tab:X-ray}
\end{table}

\begin{figure}[htb]
\includegraphics[width=0.60\textwidth]{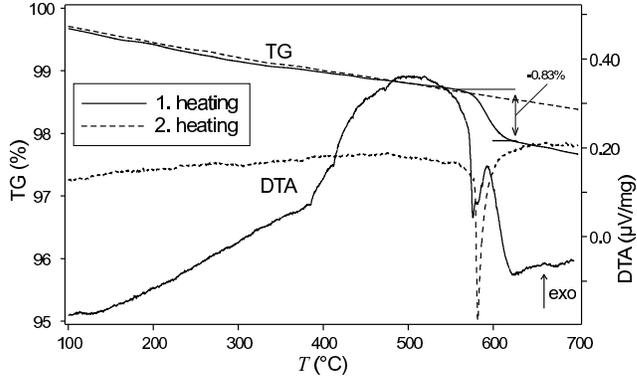}
\caption{First and second DTA heating run of an undoped K$_2$SO$_4$ crystal with 10\;K/min}
\label{fig:DTA-TG}
\end{figure}

All K$_2$SO$_4$ samples showed in the first heating run a mass loss in the order of $0.5-0.9$\% that occurs together with the PT. It must be assumed that traces of water are incorporated in the (formally anhydrous) crystal structure and that the structural changes upon the PT lead to the emanation. As this process is connected with heat exchange, the DTA peak related with the PT is superimposed by this heat (solid DTA curve in Fig. \ref{fig:DTA-TG}). In the second and all subsequent heating runs (dashed lines) no mass loss occurs and the PT peak occurs without superimposed secondary effects. Only the first DTA heating curve, before the irreversible mass loss, showed a remarkable exothermal bent starting near $400^{\,\circ}$C that is completely absent for all further DTA runs. It should be mentioned that Miyake \cite{Miyake81} observed for $T>387^{\,\circ}$C precursor phenomena for K$_2$SO$_4$ crystals being also crystallized from aqueous solution that might be related to traces of the solvent too.

Arnold et al. \cite{Arnold81} attributed precursor phenomena of the K$_2$SO$_4$ PT to OH$_3^+$ ions that are incorporated in crystals grown at room temperature from aqueous solution. In the $T$ region from 300 to $500^{\,\circ}$C (573 to 773\;K) these ions become mobile and decay finally at the PT under emanation of water. Single crystals are usually destroyed by this process. Chen and Chen \cite{Chen94} reported that crystals of K$_2$SO$_4$ burst into pieces as they were heated to $500^{\,\circ}$C and they attributed this phenomena to the OH$_3^+$ ions which reported by Arnold et al \cite{Arnold81}. It is well known that pure potassium hydrogen sulfate were 50\% of the K$^+$ ions are replaced by H$^+$, melts at $\approx200^{\,\circ}$C and decomposes upon further heating 
\begin{equation}
\text{2\;KHSO}_4 \stackrel{-\text{H}_2\text{O}}{\longrightarrow} \text{K}_2\text{S}_2\text{O}_7 \stackrel{-\text{SO}_3}{\longrightarrow} \text{K}_2\text{SO}_4
\label{eq:KHSO4}
\end{equation}
under mass loss to potassium sulfate \cite{Holleman-Wiberg56}. The theoretical mass loss that can be calculated for the first step of equation (\ref{eq:KHSO4}) is 6.6\%. The mass loss that is typically observed upon the first heating of undoped K$_2$SO$_4$ ($\approx0.83$\%, Fig. \ref{fig:DTA-TG}) is as much as $\frac{1}{8}$ of this value --- thus indicating that a considerable amount of K$^+$ ions is substituted by H$^+$ (or OH$_3^+$, respectively) in the K$_2$SO$_4$ crystal grown from aqueous solution. It should be remarked, however, that the present measurements do not allow the determination, how water is contained in the crystal: As OH$_3^+$ \cite{Arnold81,Chen94} or simply as molecule in K$_2$SO$_4\cdot x$ H$_2$O whith $x\approx0.1$. A real potassium sulfate hydrate K$_2$SO$_4\cdot n$ H$_2$O ($n$ -- integer) is not known under the current conditions.

Data for the PT of pure K$_2$SO$_4$ and for K$_2$SO$_4$:Fe$^{2+}$, K$_2$SO$_4$:Cu$^{2+}$, K$_2$SO$_4$:Cd$^{2+}$ were determined by a DTA measurement with multiple heating/cooling cycles with rates of $\pm20$, $\pm10$, $\pm5$, $\pm3$, $\pm2$, and $\pm1$\;K/min. Only samples that had already lost their water in a first DTA run were used for these DTA heating/cooling cycles with different rates. For all cycles extrapolated onsets and area of the PT peak were determined individually. Calibration measurements with Zn and BaCO$_3$ were performed and analyzed identically. Depending on heating rate, the $T$ error for the melting of Zn ranged from $-0.6$\;K to $-3.9$\;K and the sensitivity was in the range $475-560$\;$\mu$Vs/J. For BaCO$_3$ the $T$ error was $-0.5\ldots+1$\;K and the sensitivity was $358-451$\;$\mu$Vs/J.

\begin{figure}[htb]
\includegraphics[width=0.60\textwidth]{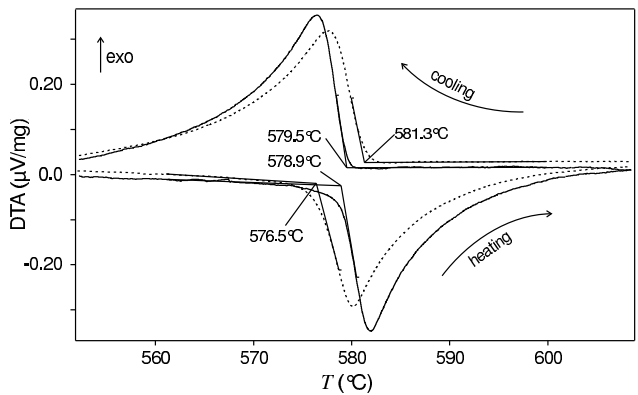}
\caption{DTA heating and cooling curves($\pm10$\;K/min) for pure K$_2$SO$_4$ (full lines) and for K$_2$SO$_4$:Fe$^{2+}$ (dashed lines)}
\label{fig:DTA_2x}
\end{figure}

Typical original DTA heating and cooling curves around the PT peak are shown for K$_2$SO$_4$ and for K$_2$SO$_4$:Fe$^{2+}$ in Fig. \ref{fig:DTA_2x}. $T_\text{t}$ can be obtained from the intersection of the extended basis line with the tangent at the inclination point (extrapolated onset). It is remarkable that for pure K$_2$SO$_4$ $T_\text{t}$ as obtained from the heating or cooling run, respectively, does not differ considerably. In contrast, the onset of the peak is lower in the heating curve and higher in the cooling curve for K$_2$SO$_4$:Fe$^{2+}$. This phenomenon will be discussed later in Fig.~\ref{fig:DTA-all}.

\begin{figure}[htb]
\includegraphics[width=0.60\textwidth]{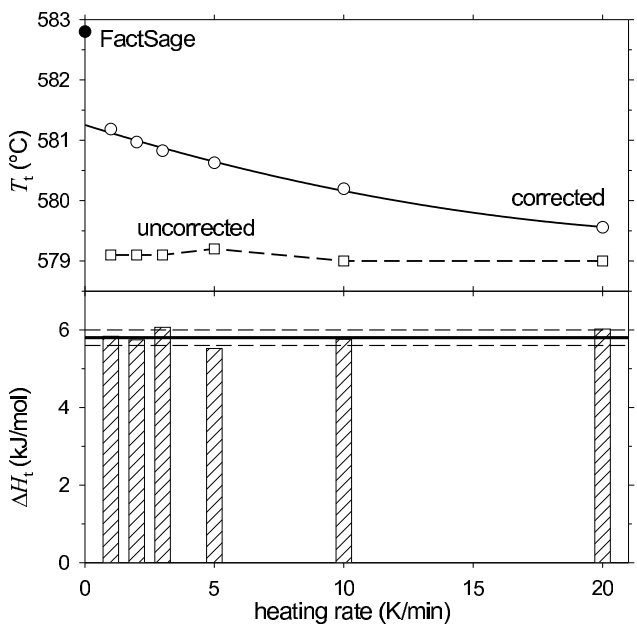}
\caption{Top: Uncorrected and corrected phase transformation temperature of pure K$_2$SO$_4$ from this work as compared to the FactSage 5.3 database value \cite{FactSage5_4_1}. Bottom: Enthalpy of the phase transformation $\it\Delta$$H_\text{t} = 5.8$\;kJ/mol}
\label{fig:PT-data}
\end{figure}

Fig. \ref{fig:PT-data} shows that the uncorrected extrapolated onsets of the PT peak for all heating rates (squares) are nearly independent on the heating rate at $579.1^{\,\circ}$C. The corrected values (circles) are obtained by adding the errors that were obtained from the interpolated Zn and BaCO$_3$ calibration measurements and can be extrapolated (for zero heating rate) to the PT temperature $T_\text{t} = 581.3^{\,\circ}$C of pure K$_2$SO$_4$. This value is by 1.5\;K lower as literature data \cite{FactSage5_4_1}. From the peak areas of the PT peaks one obtains with the sensitivity values (interpolated between Zn and BaCO$_3$) the enthalpy of the PT $\it\Delta$$H_\text{t} = (5.8\pm0.2)$\;kJ/mol. This value is smaller as the database value 8.954\;kJ/mol \cite{FactSage5_4_1} but larger than measured by Miyake et al. \cite{Miyake81} who obtained 4.28\;kJ/mol.

\begin{figure}[htb]
\includegraphics[width=0.60\textwidth]{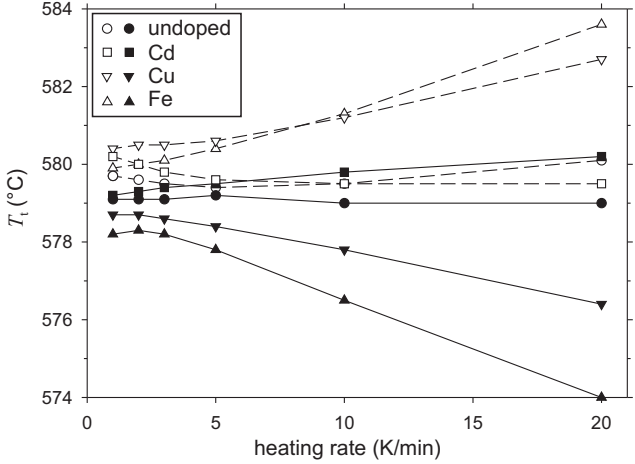}
\caption{Uncorrected extrapolated onsets of the PT peak during heating (full symbols) and cooling runs (empty symbols) for all samples and heating/cooling rates from 20\;K/min down to 1\;K/min}
\label{fig:DTA-all}
\end{figure}

Fig. \ref{fig:DTA-all} collects extrapolated onsets that were obtained for all heating/cooling rates and for all samples. Uncorrected values had to be used for this comparison, as a good correction could not be obtained for the cooling runs due to the supercooling of the BaCO$_3$ PT. Fortunately, this restriction does not influence the trends that are obvious from this figure, as any possible correction would influence every measurement data identically. It turns out that the values for K$_2$SO$_4$ and for K$_2$SO$_4$:Cd$^{2+}$ do not differ much, irrespective of the absolute rate. In contrast, for K$_2$SO$_4$:Cu$^{2+}$ and for K$_2$SO$_4$:Fe$^{2+}$ the onset temperatures are found to depend on the heating/cooling rate if the rate exceeds 3\;K/min. Actually it must be assumed that a rate dependence for lower rates is not observed for heating/cooling rates below $\dot{T}\approx\it\Delta$$T_0 /\tau$ where $\tau = 49$\;s is the time constant of the DTA set-up and $\it\Delta$$T_0 = \lim_{\dot{T}\rightarrow0}\it\Delta$$T$ is the temperature difference to be detected. If we assume $\it\Delta$$T_0=1.5$\;K (limit value for Fe and Cu doping) one arrives at $\dot{T}\approx 1.8$\;K/min. This means only for heating rates below this limit the sample is in thermal equilibrium with its surrounding. The calculated limit heating rate is in good agreement with the observed saturation for $\dot{T}\leq3$\;K/min.

The difference between extrapolated onsets in heating and cooling curves is almost vanishing for undoped K$_2$SO$_4$ (for $\dot{T}=10$\;K/min shown in Fig. \ref{fig:DTA_2x}, for all rates shown by circles in Fig. \ref{fig:DTA-all}). It is not unexpected that the K$_2$SO$_4$:Cd$^{2+}$ crystals (squares in Fig. \ref{fig:DTA-all}) showed almost the same behavior, as the Cd concentration was found to be very small (0.014\;ma\%) and cannot influence remarkably the PT. The PT peak for all heating and cooling curves of these samples behaves like it should be expected for a first order phase transformation of a pure substance. Neither precursor phenomena (except the release of water in the first heating run as described above) nor kinetic effects can be observed for these samples.

The behavior is different for the samples that were doped by Fe (0.076\;ma\%, triangles up in Fig. \ref{fig:DTA-all}) or Cu (0.192\;ma\%, triangles down in Fig. \ref{fig:DTA-all}), respectively, where a remarkable influence of $\dot{T}$ on $T_\text{t}$ can be observed. The PT peak starts at lower $T$ upon heating and starts at higher $T$ upon cooling. This temperature shift increases with the rate of temperature change and reaches almost 10\;K for K$_2$SO$_4$:Fe$^{2+}$ and 20\;K/min. Such behavior is the opposite of a normal hyteresis where one should expect just a delay of PT resulting in a higher $T_\text{t}$ upon heating and a lower $T_\text{t}$ upon cooling.

\begin{figure}[htb]
\includegraphics[width=0.30\textwidth]{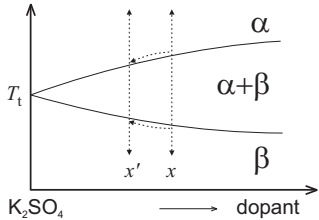}
\caption{Thermodynamic model for the explanation of the rate dependence of $T_\text{t}$}
\label{fig:Model}
\end{figure}

Instead, Fig. \ref{fig:Model} could possibly give an explanation of the observed phenomena on a thermodynamic basis: As the PT between the $\alpha$ and $\beta$ phases of K$_2$SO$_4$ is a first order phase transformation, solid solutions cannot transform at one temperature. Instead, a 2-phase region must separate the two 1-phase regions $\alpha$-K$_2$SO$_4$ and $\beta$-K$_2$SO$_4$; in analogy to the lens-shaped liquid/solid region that is typically for the melting of solid solutions. For high heating/cooling rates the phase boundaries limiting the 2-phase region are crossed close to the average composition $x$ of the solid solution. If the rate of $T$ change is sufficiently slow, segregation can take place if $T$ approaches the phase boundary, leading to a solid solution with lower concentration of the solute $x'$ and some minor amounts of the pure solvent, or of some other intermediate phase, respectively. If the composition of the K$_2$SO$_4$ solid solution is ``wandering" along the phase boundaries, $T_\text{t}$ approaches the equilibrium value of pure K$_2$SO$_4$. If this model is true, the extrapolated onsets should for very high $\dot{T}$ converge to that $T$ were the limits of the 2-phase region intersects $x$. Unfortunately, such saturation could not be observed under the current experimental conditions.

\section{Conclusions}

The DTA results presented in this study confirm, that the PT between $\beta$- and $\alpha$-K$_2$SO$_4$ is of first order. Pretransition effects below $T_\text{t}$ could be attributed to water that is included in the crystals grown from aqueous solution. Hence, previous reports on such effects \cite{Choi93,Diosa05,Miyake81} should be handled with care as it is not completely clear whether the observed effects are really intrinsic in the potassium sulfate (mobility of K$^+$ or SO$_4^{2-}$, respectively) or due to the mobility of extrinsic species like OH$_3^+$.

The kinetic properties of the $\alpha\leftrightarrow\beta$ transformation that were observed for K$_2$SO$_4$:Cu$^{2+}$ and K$_2$SO$_4$:Fe$^{2+}$ cannot be explained by a simple kinetic model based on thermally activated processes, as such processes should shift the PT peak for higher heating rate $\dot{T}$ to higher $T$, in contrast to the experimental results. Instead, a thermodynamic model is presented that can explain the observations on the basis of an extented 2-phase region between the $\beta$-K$_2$SO$_4$ and $\alpha$-K$_2$SO$_4$ 1-phase regions of K$_2$SO$_4$ based solid solutions.

\ack{Awarding a scholarship from the ``Germany Academic Exchange Service" (DAAD) for this work is gratefully acknowledged. The authors want to express their gratitude to M. Schmidbauer (IKZ Berlin) for X-ray characterization.


\end{document}